\pdfoutput=1
\documentclass[12pt]{article}
\usepackage[utf8]{inputenc}
\usepackage[height=8.85in,width=6.45in]{geometry}
\usepackage{hyperref}
\usepackage{todonotes}
\usepackage[most]{tcolorbox}

\usepackage{array}

\newtcolorbox[auto counter, number within=section]{statementbox}[1][]{
  colback=white, colframe=black,
  title=Statement~\thetcbcounter,
  fonttitle=\bfseries,
  boxrule=0.5pt,
  arc=2mm,
  left=6pt, right=6pt, top=6pt, bottom=6pt,
  #1
}

\newtcolorbox[auto counter, number within=section]{definitionbox}[1][]{
  colback=gray!10, colframe=gray!80,
  title=Definition~\thetcbcounter,
  fonttitle=\bfseries,
  boxrule=0.5pt,
  arc=2mm,
  left=6pt, right=6pt, top=6pt, bottom=6pt,
  #1
}

\newtcolorbox[auto counter, number within=section]{theorembox}[1][]{
  colback=gray!10, colframe=blue!80,
  title=Theorem~\thetcbcounter,
  fonttitle=\bfseries,
  boxrule=0.5pt,
  arc=2mm,
  left=6pt, right=6pt, top=6pt, bottom=6pt,
  #1
}

\newtcolorbox[auto counter, number within=section]{propositionbox}[1][]{
  colback=gray!10, colframe=green!60!black,
  title=Lemma~\thetcbcounter,
  fonttitle=\bfseries,
  boxrule=0.5pt,
  arc=2mm,
  left=6pt, right=6pt, top=6pt, bottom=6pt,
  #1
}

\newtcolorbox[auto counter, number within=section]{remarkbox}[1][]{
  colback=white, colframe=gray!50,
  title=Remark~\thetcbcounter,
  fonttitle=\bfseries,
  boxrule=0.3pt,
  arc=1.5mm,
  left=6pt, right=6pt, top=6pt, bottom=6pt,
  #1
}

\newtcolorbox{graybox}{breakable,
  colback=gray!40,     
  colframe=gray!15,    
  boxrule=0.2pt,       
  arc=0mm,             
  left=5pt, right=5pt, top=5pt, bottom=5pt
}

\newtcolorbox{lightgraybox}{breakable,
  colback=gray!7.5,     
  colframe=gray!15,    
  boxrule=0.2pt,       
  arc=0mm,             
  left=5pt, right=5pt, top=5pt, bottom=5pt
}

\usepackage{amsfonts}
\usepackage{amssymb}
\usepackage{amsmath}
\usepackage{amsthm}
\usepackage{mathtools}
\usepackage{cleveref}
\usepackage{url}
\usepackage{tikz}
\usepackage{tikz-cd}
\usepackage{adjustbox}
\usepackage{enumitem}
\usepackage{subcaption}
\usepackage{graphicx}
\usepackage{afterpage}
\usepackage{changepage}
\usepackage{authblk}
\usepackage{longtable}
\usepackage{bm}
\usepackage{multicol}
\usepackage{wrapfig}
\usepackage{enumitem}
\usepackage{braket}

\usetikzlibrary{knots}

\theoremstyle{plain}

\theoremstyle{definition}

\theoremstyle{remark}

\crefname{maintheorem}{Theorem}{Theorems}
\crefname{claim}{Claim}{Claims}

\newtheoremstyle{restated}
    {\topsep}{\topsep} 
    {\itshape}         
    {}                 
    {\bfseries}        
    {.}                
    { }                
    {\thmname{#1} \ref{#3} {\normalfont(Restated)}}
    
\theoremstyle{restated}
    \newtheorem{restate-theorem}{Theorem}
    \newtheorem{restate-proposition}{Proposition}
    \newtheorem{restate-corollary}{Corollary}
    
\numberwithin{equation}{section}
\numberwithin{table}{section}

\newcommand{\CC}{\mathbb{C}}

\newcommand{\MO}{\mathbb{MO}}
\newcommand{\bR}{\mathbb{R}}

\renewcommand{\arraystretch}{1.4}

\renewcommand{\epsilon}{\varepsilon}

\DeclareMathOperator{\Ima}{Im}
\DeclareMathOperator{\Ker}{Ker}

\DeclareMathOperator{\Tr}{Tr}

\newcommand{\sT}{\mathsf{T}}

\newcommand{\bZ}{\mathbb{Z}}
\newcommand{\bN}{\mathbb{N}}

\newcommand{\mbg}{\mathbf{g}}
\newcommand{\mbh}{\mathbf{h}}

\newcommand{\cA}{\mathcal{A}}

\newcommand{\cM}{\mathcal{M}}
\newcommand{\cE}{\mathcal{E}}

\def\RP{\mathbb{RP}}

\def\CC{\mathbb{CC}}
\def\MO{\mathbb{MO}}

\title{Generalization of anomaly formula for time-reversal symmetry in (2+1)D abelian bosonic TQFTs}

\author{Ippo Orii}

\affil{Kavli Institute for the Physics and Mathematics of the Universe, \\
University of Tokyo, Kashiwa, Chiba 277-8583, Japan}

\date{}
\begin{document}
    \maketitle

    \begin{abstract}
We study time-reversal symmetry in $(2+1)$D abelian bosonic topological phases. Time-reversal anomalies in such systems are classified by $\mathbb{Z}_2 \times \mathbb{Z}_2$ symmetry-protected topological (SPT) phases in $(3+1)$D, and can be diagnosed via partition functions on manifolds such as $\mathbb{RP}^4$ and $\mathbb{CP}^2$. These partition functions are related by the anomaly formula
\begin{equation*}
    Z(\mathbb{RP}^4)\, Z(\mathbb{CP}^2) = \theta_{\mathcal{M}},
\end{equation*}
where $\theta_\mathcal{M}$ is the Dehn twist phase associated with the crosscap state.

Meanwhile, the existence of gapped boundaries is constrained by so-called higher central charges $\xi_n$, which serve as computable invariants encoding obstruction data. Motivated by the known relation $Z(\mathbb{CP}^2) = \xi_1$, we propose a generalization of the anomaly formula that involves both the higher central charges $\xi_n$ and a new time-reversal invariant $\eta_n$. Introducing a distinguished subset $\mathcal{M}^n \subset \mathcal{A}$ of anyons, we establish the relation
\begin{equation*}
    \eta_n \cdot \xi_n = \frac{\sum_{a \in \mathcal{M}^n} \theta(a)^n}{\left| \sum_{a \in \mathcal{M}^n} \theta(a)^n \right|},
\end{equation*}
which generalizes the known anomaly formula.

We analyze the algebraic structure of $\mathcal{M}^n$, derive consistency relations it satisfies, and clarify its connection to the original anomaly formula. 
\end{abstract}

    \setcounter{tocdepth}{2}
    \clearpage
    \tableofcontents
    \clearpage

    \section{Introduction and Summary}
Topological quantum field theories (TQFTs) in \(2+1\) dimensions have been studied for over three decades as a fertile intersection of particle physics, topology, and category theory. In addition to their mathematical richness, they play a key role in condensed matter physics, providing effective low-energy descriptions of two-dimensional gapped phases. A canonical example is the fractional quantum Hall effect, which admits a natural description in terms of \((2+1)\)D TQFTs.

In the presence of time-reversal symmetry, such systems are subject to nontrivial consistency conditions. In particular, time-reversal anomalies in \((2+1)\)D bosonic abelian systems are classified by symmetry-protected topological (SPT) phases in \((3+1)\)D with time-reversal symmetry, which are believed to be labeled by a \(\mathbb{Z}_2 \times \mathbb{Z}_2\) classification~\cite{kapustin2014symmetryprotectedtopologicalphases}. These two \(\mathbb{Z}_2\) invariants are captured by partition functions on closed four-manifolds, specifically \(\mathbb{RP}^4\) and \(\mathbb{CP}^2\), given by
\begin{equation}
\begin{aligned}
    Z(\mathbb{RP}^4) &= \frac{1}{|\cA|^{1/2}} \sum_{a \in \Ker(1 - \sT)} \theta(a)\, \eta(a), \\
    Z(\mathbb{CP}^2) &= \frac{1}{|\cA|^{1/2}} \sum_{a \in \cA} \theta(a) = e^{2\pi i c_- / 8}, \label{Z(CP^2) in intro}
\end{aligned}
\end{equation}
where \(c_-\) denotes the chiral central charge, $\cA$ denotes the set of all anyons, \(\Ker(1 - \sT)\) denotes the set of time-reversal invariant anyons, and \(\eta(a)\) denotes local Kramers degeneracy which will be defined in~\eqref{def of eta}.

One well-known constraint is that \(c_- \equiv 0\) or \(4 \mod 8\) must hold for a bosonic topological phase to admit time-reversal symmetry. Another fundamental constraint on time-reversal anomalies is expressed as
\begin{equation}
    Z(\mathbb{RP}^4)\, Z(\mathbb{CP}^2) = \theta_{\mathcal{M}},
\end{equation}
as discussed in~\cite{Barkeshli:2016mew, Wang_2017, Orii_2025}, where \(\theta_{\mathcal{M}}\) is a phase associated with a Dehn twist around the crosscap state. Understanding and detecting time-reversal anomalies is a central theme in the study of topological phases and continues to be an active area of research.

\vspace{1em}

On the other hand, in the classification of topological orders, the existence of gapped boundaries plays a crucial role. Although determining whether a given bosonic topological order admits a gapped boundary is generally difficult, a recent refinement using the \emph{higher central charges} \(\xi_n\), labeled by a positive integer \(n\), provides a necessary and sufficient condition~\cite{Kaidi_2022}. These invariants serve as computable indicators that capture the precise obstruction to the existence of a gapped boundary.

For an abelian bosonic \((2+1)\)D topologically ordered phase, the invariant \(\xi_n\) is defined by
\begin{equation}
    \xi_n = \frac{\sum\limits_{a\in\cA}  \theta(a)^n}{\left| \sum\limits_{a\in\cA}  \theta(a)^n \right|}, \label{eq:xi_n}
\end{equation}
where the sum runs over all anyons \(a\) and \(\theta(a)\) is the topological spin of the anyon.

In particular, \(\xi_1\) gives the chiral central charge:
\begin{equation}
    \xi_1 = e^{2\pi i c_- / 8},
\end{equation}
recovering~\eqref{Z(CP^2) in intro}.

\vspace{1em}

Motivated by these observations, we propose a generalization of the anomaly formula. Let us define
\begin{equation}
    \eta_n \coloneqq \frac{\sum\limits_{a \in \Ker(1 - \mathsf{T})} \big( \theta(a)\, \eta(a) \big)^n}{\left| \sum\limits_{a \in \Ker(1 - \mathsf{T})} \big( \theta(a)\, \eta(a) \big)^n \right|}.
\end{equation}
We then consider the product \(\eta_n \cdot \xi_n\), which we find can be expressed as
\begin{equation}
    \eta_n \cdot \xi_n = \frac{\sum\limits_{a \in \mathcal{M}^n} \theta(a)^n}{\left| \sum\limits_{a \in \mathcal{M}^n} \theta(a)^n \right|},
\end{equation}
where
\begin{equation}
    \mathcal{M}^n \coloneqq \left\{ a \in \mathcal{A} \,\middle|\, B(a, n b) = \eta(n b)\ \text{for all } b \in \Ker(1 - \mathsf{T}) \right\}.
\end{equation}
Here, \(B\) is a bilinear pairing determined by the topological spin \(\theta\).

We proceed to study the algebraic structure of \(\mathcal{M}^n\) and derive constraints on its form. This analysis identifies the subset of anyons that contribute nontrivially to the higher central charge and thereby provides a refined understanding of anomaly indicators in time-reversal invariant topological phases.

\paragraph{Organization of the paper :}
In Section~2, we review the basic structure of abelian bosonic topological orders and the action of time-reversal symmetry on anyon data. In Section~3, we discuss the notion of the crosscap, which plays a crucial role in analyzing TQFTs on non-orientable manifolds. We also review our previous work~\cite{Orii_2025}, which is roughly said the case $n=1$. Our present analysis can be viewed as a direct generalization of that work, and familiarity with it may help in understanding the main results of this paper.
In Section~4, we revisit the definitions of the chiral and higher central charges, and summarize several known constraints related to gapped boundaries. Finally, in Section~5, we present our main results: a generalization of the anomaly formula and a detailed analysis of the algebraic structure of the subset \(\mathcal{M}^n\), which controls contributions to the higher central charge.

    \section{Basic of time-reversal symmetry}
\subsection{Defining data of abelian systems}
Let us define the minimal data required for an abelian anyon system. Here, we consider abelian \emph{bosonic} systems, which are well-defined without specifying a spin structure on the spacetime manifold. The required data are as follows:
\begin{lightgraybox}
    \begin{itemize}
  \item \( \mathcal{A} \): a finite abelian group\footnotemark{} of anyons (i.e., the group of topological charges).
  \item \( \theta \): the topological spin, a function \( \theta \colon \mathcal{A} \to U(1) \) which is non-degenerate, quadratic, and homogeneous.
  \item \( c_- \): the chiral central charge, an integer \( c_- \in \mathbb{Z} \) satisfying the Gauss sum constraint.
\end{itemize}
\end{lightgraybox}

\footnotetext{In general, the fusion rules of anyons are defined through the decomposition of the tensor product into a direct sum of simple objects. A theory is called \emph{Abelian} if the fusion of any two simple anyons results in another simple anyon—that is, the tensor product does not decompose further. In such cases, we use additive notation: for example, we write \( a + b \coloneqq a \otimes b \) and \( a - b \coloneqq a \otimes \overline{b} \), where \( \overline{b} \) denotes the antiparticle of \( b \).}

Given the data above, we define the braiding phase (the \( S \)-matrix entry or monodromy pairing) by:
\begin{equation}
    \begin{array}{rccc}
        B\colon & \mathcal{A} \times \mathcal{A} & \longrightarrow & U(1) \\
               & \rotatebox{90}{$\in$} & & \rotatebox{90}{$\in$} \\
               & (a, b) & \longmapsto & \theta(a + b)\theta(a)^{-1}\theta(b)^{-1}
    \end{array}.
\end{equation}

We use the following terminology:
\begin{itemize}
  \item \( \theta \) is called \emph{non-degenerate} if the associated braiding \( B \) is a non-degenerate pairing.
  \item \( \theta \) is called \emph{quadratic} if \( B \) is bihomomorphic.
  \item \( \theta \) is called \emph{homogeneous} if
  \begin{equation}
      \theta(n a) = \theta(a)^{n^2} \quad \text{for all } a \in \mathcal{A},\, n \in \mathbb{Z}.
  \end{equation}
\end{itemize}

Finally, the Gauss sum constraint is given by:
\begin{equation}
    \frac{1}{|\mathcal{A}|^{1/2}} \sum_{a \in \mathcal{A}} \theta(a) = e^{2\pi i c_- / 8}.
\end{equation}

\subsection{Time-reversal action}
Let us now encode the time-reversal action on anyons in our setup. We define the operator \( \mathsf{T} \colon \mathcal{A} \to \mathcal{A} \) by the condition:
\begin{equation}
    \mathsf{T} \colon \mathcal{A} \to \mathcal{A}, \quad \text{such that }\sT^2=\mathrm{id}_\cA,\quad \theta(\mathsf{T}a)\, \theta(a) = 1 \quad \text{for all } a \in \mathcal{A}.
\end{equation}
The condition \( \theta(\mathsf{T}a)\, \theta(a) = 1 \) captures the anti-unitary nature of time-reversal symmetry (see, e.g., \cite{Barkeshli:2016mew}), and we will sometimes write \(\sT^2=\mathrm{id}_\cA\) compactly as \( \mathsf{T}^2 = 1 \). Using this definition, we can derive several relations involving the braiding phase \( B \):

\begin{lightgraybox}
\begin{equation}
    \begin{aligned}
        B(a, b) &= B(\mathsf{T}a, \mathsf{T}b)^{-1}, \\
        B(\mathsf{T}a, b) &= B(a, \mathsf{T}b)^{-1}, \\
        B\big((1 - \mathsf{T})a, b\big) &= B\big(a, (1 + \mathsf{T})b\big).
    \end{aligned}
\end{equation}
\end{lightgraybox}

These follow directly from the definition of \( B \) and $\theta$.

\subsection{Local Kramers degeneracy}
\label{H3}
\paragraph{Definition of Local Kramers degeneracy :}We now encode the notion of local Kramers degeneracy, which is often interpreted as the local eigenvalue of \( \mathsf{T}^2 \). We define a function
\begin{equation}
    \eta(a) = 
    \begin{cases}
        \pm1 & \text{if } a \in \Ker(1 - \mathsf{T}), \\
        0 & \text{otherwise},
    \end{cases}
    \label{def of eta}
\end{equation}
satisfying the following conditions:
\begin{equation}
    \eta(a)\, \eta(b) = \eta(a + b) \quad \text{for all } a, b \in \Ker(1 - \mathsf{T}).
\end{equation}
By definition, we immediately see that \( \eta(0) = 1 \), since \( \eta(0 + 0) = \eta(0)^2 \). In addition, we easily obtain
\begin{equation}
    \eta(a)=\eta(-a)=\eta(\sT a).
\end{equation}

\paragraph{Symmetry fractionalization :} Local Kramers degeneracy, as introduced above, can be understood as a special case of symmetry fractionalization. Symmetry fractionalization arises when we attempt to define how a global symmetry group \( G \) acts locally on individual anyons in a topological phase. In general, it is not always possible to consistently localize the symmetry action on each anyon—this obstruction is referred to as a \(H^3\) obstruction or a symmetry localization anomaly (see~\cite{Barkeshli_2019} for a precise discussion). When this obstruction is present, the structures of symmetry are described by a 2-group~\cite{Tachikawa_2020, Benini_2019}, a kind of extended notion of a group by a one-form symmetry. When this obstruction is absent, the global symmetry can be consistently decomposed into local actions on anyons, leading to a well-defined notion of symmetry fractionalization. Also note that in the abelian cases, \(H^3\) obstruction for time-reversal was shown to vanish~\cite{Orii:2025hgn}.
 
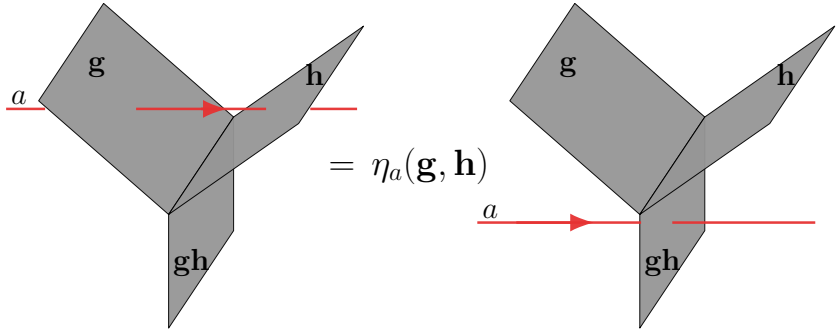
\begin{figure}[t]
    \centering
    \scalebox{0.8}{ 
\begin{tikzpicture}[x={(0.8cm,0cm)}, y={(0.4cm,0.6cm)}, z={(0cm,0.7cm)},
  line join=round, scale=2.7]

  \definecolor{mygray}{RGB}{150,150,150}
\definecolor{myred}{RGB}{230,50,50}

\tikzstyle{sheet} = [fill=mygray,opacity=0.95,draw=black,very thin]
\tikzstyle{byline} = [draw=myred, fill=myred, line width=1.2pt, opacity=0.95]

  \begin{scope}
    \draw[sheet] (0,0,0) -- (0,0,-1) -- (0,1,-1) -- (0,1,0) -- cycle; 
    \draw[sheet] (-1,0,1) -- (-1,1,1) -- (0,1,0) -- (0,0,0) -- cycle; 
    \draw[sheet] (0,1,0) -- (1,1,0.8) -- (1,0,0.8) -- (0,0,0) -- cycle; 

    \draw[byline] (-1.5,0.5,0.5) -- (-1.2,0.5,0.5);
    \draw[byline] (-0.5,0.5,0.5) -- (0.5,0.5,0.5);
    \draw[byline] (0.84,0.5,0.5) -- (1.2,0.5,0.5);

    \node at (-0.9,0.7,0.7) {\large$\mathbf{g}$};
    \node at (1.05,0.15,1.1) {\large$\mathbf{h}$};
    \node at (0.1,0.15,-0.55) {\large$\mathbf{gh}$};
    \node at (-1.4,0.5,0.6) {\large$a$};

    \draw[-{Latex[length=4mm]}, byline] (0,0.5,0.5) -- (0.2,0.5,0.5);
  \end{scope}

  \node at (1.6,0.5,0) {\Large$=\, \eta_a(\mathbf{g},\mathbf{h})$};

  \begin{scope}[xshift=2.5cm]
    \draw[sheet] (0.5,0,0) -- (0.5,0,-1) -- (0.5,1,-1) -- (0.5,1,0) -- cycle;
    \draw[sheet] (-0.5,0,1) -- (-0.5,1,1) -- (0.5,1,0) -- (0.5,0,0) -- cycle;
    \draw[sheet] (0.5,1,0) -- (1.5,1,0.8) -- (1.5,0,0.8) -- (0.5,0,0) -- cycle;

    \draw[byline] (-1.0,0.5,-0.5) -- (0.26,0.5,-0.5);
    \draw[byline] (0.5,0.5,-0.5) -- (1.6,0.5,-0.5);

    \node at (-0.4,0.7,0.7) {\large$\mathbf{g}$};
    \node at (1.55,0.15,1.1) {\large$\mathbf{h}$};
    \node at (0.6,0.15,-0.55) {\large$\mathbf{gh}$};
    \node at (-0.9,0.5,-0.4) {\large$a$};

    \draw[-{Latex[length=4mm]}, byline] 
    (-0.7,0.5,-0.5) -- (-0.1,0.5,-0.5);
  \end{scope}

\end{tikzpicture}
}

    \caption{Symmetry fractionalization: each gray sheet represents a domain wall labeled by a group element, and an anyon is acted upon by that element when passing through the wall.}

    \label{fig:sym frac}
\end{figure}

The quantities \( \eta_a(\mbg, \mbh) \) encode the phase difference between the sequential local actions of \( \mbg \) and \( \mbh \) on an anyon \( a \), and the local action of their product \( \mbg \mbh \). They reflect the projective nature of the localized symmetry action on the anyon (see~Figure~\ref{fig:sym frac}).

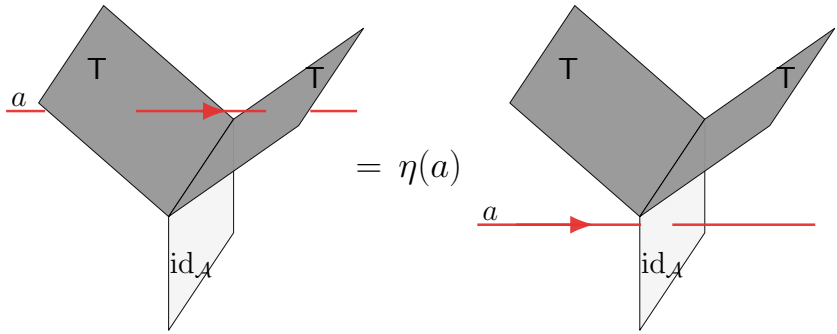
\begin{figure}[t]
    \centering
    \scalebox{0.8}{ 
\begin{tikzpicture}[x={(0.8cm,0cm)}, y={(0.4cm,0.6cm)}, z={(0cm,0.7cm)},
  line join=round, scale=2.7]

  \definecolor{mygray}{RGB}{150,150,150}
\definecolor{myred}{RGB}{230,50,50}
\definecolor{mywhite}{RGB}{245,245,245}

\tikzstyle{sheet} = [fill=mygray,opacity=0.95,draw=black,very thin]
\tikzstyle{sheetW} = [fill=mywhite,opacity=0.95,draw=black,very thin]
\tikzstyle{byline} = [draw=myred, fill=myred, line width=1.2pt, opacity=0.95]

  \begin{scope}
    \draw[sheetW] (0,0,0) -- (0,0,-1) -- (0,1,-1) -- (0,1,0) -- cycle; 
    \draw[sheet] (-1,0,1) -- (-1,1,1) -- (0,1,0) -- (0,0,0) -- cycle; 
    \draw[sheet] (0,1,0) -- (1,1,0.8) -- (1,0,0.8) -- (0,0,0) -- cycle; 

    \draw[byline] (-1.5,0.5,0.5) -- (-1.2,0.5,0.5);
    \draw[byline] (-0.5,0.5,0.5) -- (0.5,0.5,0.5);
    \draw[byline] (0.84,0.5,0.5) -- (1.2,0.5,0.5);

    \node at (-0.9,0.7,0.7) {\large$\mathsf{T}$};
    \node at (1.05,0.15,1.1) {\large$\mathsf{T}$};
    \node at (0.1,0.15,-0.55) {\large$\mathrm{id}_\cA$};
    \node at (-1.4,0.5,0.6) {\large$a$};

    \draw[-{Latex[length=4mm]}, byline] (0,0.5,0.5) -- (0.2,0.5,0.5);
  \end{scope}

  \node at (1.6,0.5,0) {\Large$=\, \eta(a)$};

  \begin{scope}[xshift=2.5cm]
    \draw[sheetW] (0.5,0,0) -- (0.5,0,-1) -- (0.5,1,-1) -- (0.5,1,0) -- cycle;
    \draw[sheet] (-0.5,0,1) -- (-0.5,1,1) -- (0.5,1,0) -- (0.5,0,0) -- cycle;
    \draw[sheet] (0.5,1,0) -- (1.5,1,0.8) -- (1.5,0,0.8) -- (0.5,0,0) -- cycle;

    \draw[byline] (-1.0,0.5,-0.5) -- (0.26,0.5,-0.5);
    \draw[byline] (0.5,0.5,-0.5) -- (1.6,0.5,-0.5);

    \node at (-0.4,0.7,0.7) {\large$\mathsf{T}$};
    \node at (1.55,0.15,1.1) {\large$\mathsf{T}$};
    \node at (0.6,0.15,-0.55) {\large$\mathrm{id}_\cA$};
    \node at (-0.9,0.5,-0.4) {\large$a$};

    \draw[-{Latex[length=4mm]}, byline] 
    (-0.7,0.5,-0.5) -- (-0.1,0.5,-0.5);
  \end{scope}

\end{tikzpicture}
}

\caption{Local Kramers degeneracy: even when an anyon is invariant under time-reversal symmetry, a nontrivial phase may appear upon acting with $\sT$ twice.}

    \label{fig:loc kram}
\end{figure}

Setting \( \mbg = \mbh = \sT \), and $a$ to be time-reversal invariant, (i.e. $a\in\Ker(1-\sT)$,) the resulting phase becomes \( \eta_a(\sT, \sT) \), which we identify with \( \eta(a) \) (see~Figure~\ref{fig:loc kram}).

    \section{Analysis of Crosscap}
\subsection{Crosscap state}
The quantity \( \eta(a) \) also admits another interpretation. When analyzing theories with time-reversal symmetry, it is common to study the so-called \emph{crosscap state}, defined as
\begin{equation}
    \ket{\CC} = Z(\MO_A \times S^1_B),
\end{equation}
where
\begin{equation}
    \MO_A = \left\{ (x, \theta) \in [-1,1] \times \bR \,\middle|\, (x, \theta) \sim (-x, \theta + \pi) \right\}.
    \label{moa}
\end{equation}
Here, we introduce subscripts \( A \) and \( B \) to distinguish the two \( S^1 \) directions. The boundary of \( \MO_A \times S^1_B \) is then
\begin{equation}
    \partial(\MO_A \times S^1_B) = S^1_A \times S^1_B = T^2.
\end{equation}
Thus, the crosscap state \( \ket{\CC} \) can be expanded in terms of a basis of the Hilbert space \( V(T^2) \). If a complete orthonormal basis of \( V(T^2) \) is given by \( \{ \ket{a} \}_{a \in \cA} \) and their modular \( S \)-transformed states \( \{ S \ket{a} \}_{a \in \cA} \),\footnote{See~\cite{Tachikawa_2017} for the construction of $V(T^2)$.} then we can write
\begin{equation}
    \begin{aligned}
        \ket{\CC} &= \sum_{a \in \cA} M_a \ket{a} \\
        &= \sum_{a \in \cA} \eta(a) S \ket{a},\label{def of crosscap state}
    \end{aligned}
\end{equation}
where \( M_a \) and \( \eta(a) \) are expansion coefficients.
Let us closely look at each coefficient.
\paragraph{$M_a$ :}From equation~\eqref{def of crosscap state}, we have
\begin{equation}
    M_a=\sum_{b\in\cA}S_{ab}\eta(b)
\end{equation}
where $S_{ab}\coloneqq\bra{a}S\ket{b}$. At the same time, a geometrical analysis shows us that
\begin{equation}
    M_a=Z(\RP^2(a)\times S^1)=\dim\left( V\big( \mathbb{RP}^2(a) \big) \right)
\end{equation}
where $\RP^2(a)$ is a real projective plane with a puncture lablled by $a\in\cA$. In \cite{Orii_2025}, we prove that this value $M_a$ is a non-negative integer for all $a\in\cA$,\footnote{This statement might appear rather trivial, since \( M_a \) denotes the dimension of a Hilbert space and is therefore expected to be an integer. However, this integrality can fail when time-reversal symmetry is not properly encoded in the theory due to the \(H^3\) obstruction which we mentioned in~\ref{H3}. But computing such obstructions explicitly is often difficult. Therefore, the fact that \( M_a \in \mathbb{Z} \) can be regarded as a necessary consistency condition for the proper implementation of time-reversal symmetry, and is thus nontrivial.
 See e.g.~\cite{Barkeshli_2019} for general discussions on encoding symmetry. See also~\cite{Orii:2025hgn} for the proof of the vanishing of the obstruction in abelian systems.} and define $\cM$ as a subset in $\cA$ satisfying
\begin{equation}
    M_a\ne0\quad\text{iff }a\in\cM. 
\end{equation}
We will review several facts associated with $\cM$ in Sec.~3.2.

\paragraph{\(\eta(a)\) :} From equation~\eqref{def of crosscap state}, we have
\begin{equation}
    \eta(a) = \bra{a} S^{-1} \ket{\CC}.
\end{equation}
On the other hand, geometric analysis provides the following expression:
\begin{equation}
    \eta(a) = \Tr_{V(S^2(a, -\mathsf{T}a))}(P_{S^2}).
\end{equation}
Here, \( V(S^2(a, a')) \) denotes the Hilbert space on the two-sphere \( S^2 \) with anyons \( a \) and \( a' \) inserted at the north and south poles, respectively. The operator \( P_{S^2} \colon V(S^2(a, a')) \to V(S^2(-\mathsf{T}a', -\mathsf{T}a)) \) implements the reflection \( (n_x, n_y, n_z) \mapsto (-n_x, n_y, n_z) \) on the spatial sphere, where
\begin{equation}
        S^2=\{\vec{n}=(n_x,n_y,n_z)\mid|\vec{n}|=1\}
\end{equation}

Setting \( a' = -\mathsf{T}a \), \( P_{S^2} \) becomes an endomorphism
\begin{equation}
    P_{S^2} \colon V(S^2(a, -\mathsf{T}a)) \to V(S^2(a, -\mathsf{T}a)).
\end{equation}
This expression appears in~\cite{Barkeshli:2016mew, Barkeshli:2017rzd, Tachikawa_2017} and is interpreted as the local Kramers degeneracy. A brief inspection shows that it is nonzero only when \( a = \mathsf{T}a \), i.e., \( a \in \Ker(1 - \mathsf{T}) \), and it takes values in \( \{\pm 1\} \), which are properties we define in~\eqref{def of eta}.\footnote{There is a subtlety here. The restriction to \( \mathbb{Z}_2 \)-valued phases is often assumed or justified based on other physical or mathematical conditions; see~\cite{Orii_2025} for further discussion.}

\begin{figure}[t]
    \centering
    \scalebox{0.8}{ 
\begin{tikzpicture}[x={(0.8cm,0cm)}, y={(0.4cm,0.6cm)}, z={(0cm,0.7cm)},
  line join=round, scale=2.7]

  \definecolor{mygray}{RGB}{150,150,150}
\definecolor{myred}{RGB}{230,50,50}
\definecolor{mywhite}{RGB}{245,245,245}
\definecolor{myblue}{RGB}{50,50,230}

\tikzstyle{sheet} = [fill=mygray,opacity=0.95,draw=black,very thin]
\tikzstyle{sheetW} = [fill=mywhite,opacity=0.95,draw=black,very thin]
\tikzstyle{byline} = [draw=myred, fill=myred, line width=1.2pt, opacity=0.95]
\tikzstyle{trapline} = [draw=myblue, fill=myblue, line width=1.2pt, opacity=0.95]

  \begin{scope}
    \draw[sheetW] (0,0,0) -- (0,0,-1) -- (0,1,-1) -- (0,1,0) -- cycle; 
    \draw[sheet] (-1,0,1) -- (-1,1,1) -- (0,1,0) -- (0,0,0) -- cycle; 
    \draw[sheet] (0,1,0) -- (1,1,0.8) -- (1,0,0.8) -- (0,0,0) -- cycle; 

    \draw[trapline] (0,0,0) -- (0,1,0);
    \node at (-0.1,0,-0.2) {\large$b$};
    \draw[-{Latex[length=4mm]}, trapline] (0,0.6,0) -- (0,0.4,0);

    \draw[byline] (-1.5,0.5,0.5) -- (-1.2,0.5,0.5);
    \draw[byline] (-0.5,0.5,0.5) -- (0.5,0.5,0.5);
    \draw[byline] (0.84,0.5,0.5) -- (1.2,0.5,0.5);

    \node at (-0.9,0.7,0.7) {\large$\mathsf{T}$};
    \node at (1.05,0.15,1.1) {\large$\mathsf{T}$};
    \node at (0.1,0.15,-0.55) {\large$\mathrm{id}_\cA$};
    \node at (-1.4,0.5,0.6) {\large$a$};

    \draw[-{Latex[length=4mm]}, byline] (0,0.5,0.5) -- (0.2,0.5,0.5);
  \end{scope}

  \node at (1.6,0.5,0) {\Large$=\, B(a,b)$};

  \begin{scope}[xshift=2.5cm]
    \draw[sheetW] (0.5,0,0) -- (0.5,0,-1) -- (0.5,1,-1) -- (0.5,1,0) -- cycle;
    \draw[sheet] (-0.5,0,1) -- (-0.5,1,1) -- (0.5,1,0) -- (0.5,0,0) -- cycle;
    \draw[sheet] (0.5,1,0) -- (1.5,1,0.8) -- (1.5,0,0.8) -- (0.5,0,0) -- cycle;

    \draw[trapline] (0.5,0,0) -- (0.5,1,0);
\node at (0.4,0,-0.2) {\large$b$};
\draw[-{Latex[length=4mm]}, trapline] (0.5,0.6,0) -- (0.5,0.4,0);

    \draw[byline] (-1.0,0.5,-0.5) -- (0.26,0.5,-0.5);
    \draw[byline] (0.5,0.5,-0.5) -- (1.6,0.5,-0.5);

    \node at (-0.4,0.7,0.7) {\large$\mathsf{T}$};
    \node at (1.55,0.15,1.1) {\large$\mathsf{T}$};
    \node at (0.6,0.15,-0.55) {\large$\mathrm{id}_\cA$};
    \node at (-0.9,0.5,-0.4) {\large$a$};

    \draw[-{Latex[length=4mm]}, byline] 
    (-0.7,0.5,-0.5) -- (-0.1,0.5,-0.5);
  \end{scope}

\end{tikzpicture}
}

    \caption{Trapped anyon: The phase associated with local Kramers degeneracy can be interpreted as the mutual braiding between an anyon and another anyon trapped at the intersection of two time-reversal domain walls.}

    \label{fig:loc kram with B}
\end{figure}
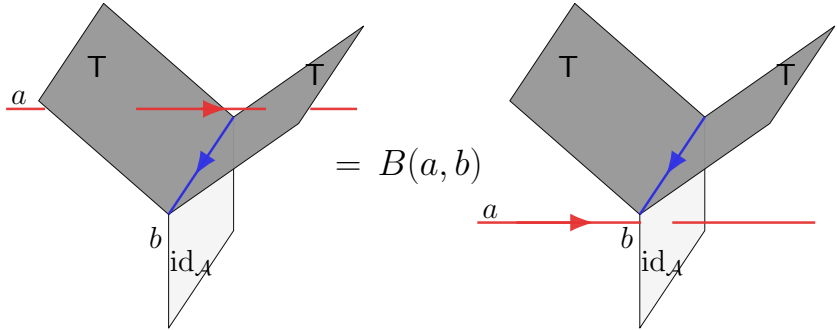

\subsection{Constraints on $M_a$ and $\mathcal{M}$}
In our previous work \cite{Orii_2025}, we defined \( M_a \) and \( \mathcal{M} \) as follows:\footnote{Here, we replace \( \mathcal{D} \) in the previous paper with \( \mathcal{M} \), since \( \mathcal{D} \) often denotes the total quantum dimension.}
\begin{equation}
    M_a=\frac{1}{|\cA|^{1/2}}\sum_{b\in\cA}B(-a,b)\eta(b),\label{def of M_a}
\end{equation}
\begin{equation}
    \cM=\{a\in\cA\mid M_a\ne0\}.\label{def of M}
\end{equation}
Then we find that each of these takes the following form:
\begin{equation}
    M_a=\begin{cases}
2^m & a\in\cM\quad\text{for some }m\in\bZ,\\
0 & a\notin\cM
\end{cases},
\end{equation}
and

\begin{equation}
    \cM=\{a\in\cA\mid  B(a,b)=\eta(b)\quad\text{for all }b\in\Ker(1-\sT)\}.
\end{equation}
From the above form of \(\cM\), one can show that it is an \(\Ima(1+\sT)\text{-torsor}\).
In addition, there is a constraint:
\begin{equation}
    2a\in\Ima(1+\sT)\quad\text{for all }a\in\cM.
\end{equation}
We derive the form of \( \mathcal{M} \) and this constraint algebraically in \cite{Orii_2025}, but we find that it can also be interpreted pictorically from the viewpoint of symmetry fractionalization.\footnote{This kind of discussion of symmetry fractionalizatioin is well-established in e.g.~\cite{Tata_2022}.}

Recall the characterization of \( \eta \) illustrated in Figure~\ref{fig:loc kram}. In such a situation, it is natural to interpret the intersection of the domain walls as trapping an anyon, as depicted in Figure~\ref{fig:loc kram with B}. The phase \( \eta(a) \) then acquires a physical interpretation: it arises as a mutual braiding between the probe anyon and the trapped anyon.

Comparing the two expressions for this phase, we arrive at the identification
\begin{graybox}
\vspace{-1.5em}
\begin{equation}
    \{ \text{trapped anyons} \}
    = \left\{ a \in \cA \mid B(a,b) = \eta(b) \quad \text{for all } b \in \Ker(1 - \sT) \right\}
    = \cM.
\end{equation}
\end{graybox}

Let us now try to guess the structure of \( \{ \text{trapped anyons} \} \), assuming we do not yet know the algebraic structure of \( \cM \). For this, we use the identity (see Appendix~\ref{App 1}):
\begin{lightgraybox}
\begin{equation}
    \Ker(1 - \sT) = \left[ \Ima(1 + \sT) \right]^\perp.
\end{equation}
\end{lightgraybox}

First, note that
\begin{equation}
    a + (1 + \sT)a' \in \{ \text{trapped anyons} \}
    \quad \text{for all } a \in \{ \text{trapped anyons} \},
\end{equation}
since
\begin{equation}
    B(a + (1 + \sT)a', b) = B(a, b)\, B((1 + \sT)a', b) = \eta(b)
    \quad \text{for all } b \in \Ker(1 - \sT).
\end{equation}
This shows that the set of trapped anyons is stable under shifts by \( \Ima(1 + \sT) \), implying a torsor structure over \( \Ima(1 + \sT) \).

Furthermore, since \( \eta(2b) = \eta(b)^2 = 1 \), we compute
\begin{equation}
    B(a, 2b) = B(2a, b) = 1 \quad \text{for all } b \in \Ker(1 - \sT),
\end{equation}
which implies
\begin{equation}
    2a \in \Ima(1 + \sT) \quad \text{for all } a \in \{ \text{trapped anyons} \}.
\end{equation}

This torsor structure and doubling constraint are exactly the properties we derived algebraically in our previous work. Thus, this physical picture provides support for our prior interpretation.

Before moving on to the next section, let us recall a condition associated with the topological spin of anyons in \( \mathcal{M} \) \cite{Barkeshli:2016mew, Orii_2025, Tachikawa_2017}, which states that
\begin{equation}
    \theta_\mathcal{M} \coloneqq \theta(a) = \text{const.} \in \{ \pm 1 \} \quad \text{for all } a \in \mathcal{M}.
\end{equation}
This is considered one of the constraints that must be satisfied for the theory to be anomaly-free \cite{Barkeshli:2017rzd}.


    \section{Gapped boundary}
\subsection{Lagrangian subgroup}
Strictly adhering to the axioms, one typically associates Hilbert spaces only to closed two-dimensional surfaces. Within this formalism, incorporating spacelike boundaries is subtle, as such boundaries do not naturally fit into the standard TQFT framework. Nonetheless, it is both natural and physically relevant to consider systems with spacelike boundaries and to investigate their properties. When the boundary itself admits a topological description---that is, when no gapless degrees of freedom appear along the boundary---we refer to it as a \emph{gapped boundary}.

Determining whether a given topological phase admits a gapped boundary is one of central questions in the study of topological phases. For abelian systems, a necessary and sufficient condition for the existence of a gapped boundary is known~\cite{Kapustin_2011,Fuchs_2013}:

\begin{lightgraybox}
\begin{center}
    A gapped boundary exists if and only if there exists a Lagrangian subgroup $L \subset \mathcal{A}$.
\end{center}
\end{lightgraybox}

Here, a \emph{Lagrangian subgroup} is defined as follows:

\begin{lightgraybox}
A subgroup $L \subset \mathcal{A}$ satisfying:
\begin{itemize}
    \item $\theta(a) = 1$ \quad for all $a \in L$,
    \item $B(a,b) = 1$ \quad for all $a, b \in L$,
    \item $|L|^2 = |\mathcal{A}|$,
    \item For every $a \in \mathcal{A} \setminus L$, there exists at least one $b \in L$ such that $B(a,b) \neq 1$.
\end{itemize}
\end{lightgraybox}
Roughly speaking, a gapped boundary can be viewed as an interface between a nontrivial TQFT and a trivial TQFT. A Lagrangian subgroup is a set of anyons whose condensation effectively trivializes the theory. Therefore, if the anyon system admits such a subgroup, one can obtain a gapped boundary by condensing the corresponding anyons in a subregion of spacetime.

Thus, verifying the existence of a gapped boundary reduces to identifying a Lagrangian subgroup. However, finding such a subgroup is generally nontrivial. Therefore, it is often useful to employ computable invariants as practical criteria for detecting the existence of gapped boundaries as we introduce in the next subsections.

\paragraph{Case when $|\cA|$ is odd :}In~\cite{Lee:2018eqa}, it was shown that any abelian time-reversal symmetric systems with odd $|\cA|$ can be viewed as a gauge theory. Therefore, it obviously admits a gapped boundary. 
 Although it lies somewhat outside the main scope of this paper, we include it here for reference, as it may be useful in future applications.

 We will show the following.
\begin{graybox}
   \begin{center}
       $\cM$ is a lagrangian subgroup for such TQFTs.
   \end{center}
\end{graybox}

Before proceeding, we establish the following proposition:
\begin{graybox}
\begin{equation}
    \Ker(1 - \mathsf{T}) = \mathrm{Im}(1 + \mathsf{T}) \quad \text{when } |\mathcal{A}| \text{ is odd}.
\end{equation}
\end{graybox}

The inclusion 
\begin{equation}
    \mathrm{Im}(1 + \mathsf{T}) \subset \Ker(1 - \mathsf{T})
\end{equation}
is immediate. Now pick any \( a \in \Ker(1 - \mathsf{T}) \). Since \( |\mathcal{A}| \) is odd, division by two is well-defined, and we have:
\begin{equation}
\begin{aligned}
    a &= \frac{a}{2} + \frac{a}{2} \\
      &= \frac{a}{2} + \frac{\mathsf{T}a}{2} \\
      &= (1 + \mathsf{T})\frac{a}{2} \in \mathrm{Im}(1 + \mathsf{T}),
\end{aligned}
\end{equation}
which shows
\begin{equation}
\Ker(1 - \mathsf{T}) \subset \mathrm{}(1 + \mathsf{T}).
\end{equation}
Therefore, we conclude
\begin{equation}
\Ker(1 - \mathsf{T}) = \mathrm{Im}(1 + \mathsf{T}).
\end{equation}

\vspace{1em}
Next we will derive the following:
\begin{graybox}
    \begin{equation}
    \mathcal{M} = \mathrm{Im}(1 + \mathsf{T}) \quad \text{when } |\mathcal{A}| \text{ is odd}.\label{M = Im}
\end{equation} 
\end{graybox}
To do so, we will use the following facts:
\begin{lightgraybox}
    \begin{itemize}
    \item When \( |\mathcal{A}| \) is odd, one can always divide any anyon by two in the following sense: for any \( a \in \mathcal{A} \), there exists a unique \( b \in \mathcal{A} \) such that \( a = 2b \). We denote this element by \( a/2 \). Note that \( \mathsf{T}(a/2) = \mathsf{T}(a)/2 \).
    
    \item The following orthogonality relation holds (see Appendix~\ref{App 1}):
    \begin{equation}
        \Ker(1 - \mathsf{T}) = \left[ \mathrm{Im}(1 + \mathsf{T}) \right]^\perp.
    \end{equation}
\end{itemize}
\end{lightgraybox}

At the beginning, it is easy to show that \( \eta(a) = 1 \) for all \( a \in \Ker(1 - \mathsf{T}) \), since
\begin{equation}
    \eta(a) = \eta\left(\frac{a}{2} + \frac{a}{2}\right) = \eta\left(\frac{a}{2}\right)^2 = 1,
\end{equation}
where we used the facts that \( a/2 \in \Ker(1 - \mathsf{T}) \) if \( a \in \Ker(1 - \mathsf{T}) \), that \( \eta \) is a homomorphism, and that \( \eta(a) \in \{ \pm 1 \} \). 
Then, for \( a \in \mathcal{M} \),
\begin{equation}
    B(a, b) = \eta(b) = 1 \quad \text{for all } b \in \Ker(1 - \mathsf{T}).
\end{equation}
Using the fact that
\begin{equation}
\Ker(1 - \mathsf{T}) = \left[ \mathrm{Im}(1 + \mathsf{T}) \right]^\perp,
\end{equation}
we obtain
\begin{equation}
\mathcal{M} \subset \mathrm{Im}(1 + \mathsf{T}).
\end{equation}
Conversely, for \( a \in \mathrm{Im}(1 + \mathsf{T}) \), the orthogonality condition implies
\begin{equation}
\mathrm{Im}(1 + \mathsf{T}) \subset \mathcal{M}.
\end{equation}
Therefore, we conclude that
\begin{equation}
\mathcal{M} = \mathrm{Im}(1 + \mathsf{T}).
\end{equation}
\vspace{1em}
Then, we easily see that
\begin{equation}
\theta_{\cM}=1
\end{equation}
since for any \( a \in \mathcal{A} \), we have
\begin{equation}
    \begin{aligned}
         \theta_{\mathcal{M}} &= \theta((1 + \mathsf{T})a) \\
    &= B(a, \mathsf{T}a)\, \theta(a)\, \theta(\mathsf{T}a) \\
    &= \eta((1 + \mathsf{T})a)^{-1} \\
    &= 1.
    \end{aligned}
\end{equation}
Here, we use the consistency condition in Appendix~\ref{App2}.

Now, define
\begin{equation}
L := \mathrm{Im}(1 + \mathsf{T}) = \Ker(1 - \mathsf{T}) = \mathcal{M}.
\end{equation}
Then, the first and second conditions for the lagrangian subgroup follow immediately from \( \theta_{\mathcal{M}} = 1 \) and the orthogonality \( \Ker(1 - \mathsf{T}) \perp \mathrm{Im}(1 - \mathsf{T}) \).

The third condition follows from the identity
\begin{equation}
|\mathcal{A}| = |\Ker(1 - \mathsf{T})| \cdot |\mathrm{Im}(1 - \mathsf{T})|.
\end{equation}

As for the fourth condition: for all \( a \in \mathcal{A} \setminus \mathrm{Im}(1 + \mathsf{T}) = \mathcal{A} \setminus L \), there exists \( b \in L = \Ker(1 - \mathsf{T}) \) such that \( B(a, b) \ne 1 \). Otherwise, if \( B(a, b) = 1 \) for all \( b \in \Ker(1 - \mathsf{T}) \), it would imply \( a \in \Ker(1 - \mathsf{T})^\perp = \mathrm{Im}(1 + \mathsf{T}) \), contradicting \( a \notin \mathrm{Im}(1 + \mathsf{T}) \).

This completes the verification.

\subsection{Chiral central charge and Higher central charge}
We have already introduced the chiral central charge as part of the input data for abelian systems. It is also known to serve as an obstruction to the existence of a gapped boundary:
\begin{lightgraybox}
\begin{center}
    A necessary condition for the existence of a gapped boundary is that
    \begin{equation}
        c_- \equiv 0 \pmod{8}.
    \end{equation}
\end{center}
\end{lightgraybox}

In the presence of time-reversal symmetry, it is known that \( c_- \) must take values in \( 0 \) or \( 4 \pmod{8} \). This can be seen as follows:
\begin{equation}
    \begin{aligned}
         e^{2\pi i c_- / 8}
    &= \frac{1}{|\mathcal{A}|^{1/2}} \sum_{a \in \mathcal{A}} \theta(a) \\
    &= \frac{1}{|\mathcal{A}|^{1/2}} \sum_{a \in \mathcal{A}} \theta(\sT a) \\
    &= \frac{1}{|\mathcal{A}|^{1/2}} \sum_{a \in \mathcal{A}} \overline{\theta(a)} \\
    &= e^{-2\pi i c_- / 8},
    \end{aligned}
\end{equation}
which implies that
\begin{equation}
    c_- \equiv 0 \text{ or } 4 \pmod{8}.
\end{equation}

This is a well-known constraint for the existence of time-reversal symmetry in bosonic systems.\footnote{For a detailed discussion of the chiral central charge in time-reversal invariant systems, see~\cite{Kobayashi_2021}.}

In particular, when \( |\cA| \) is odd, it is evident that
\begin{equation}
    c_- \equiv 0 \pmod{8}
\end{equation}
since such theories can be realized as gauge theories~\cite{Lee:2018eqa}.
However, the condition above is \emph{necessary}, but not \emph{sufficient}. Recently, a new computable invariant called the \emph{Higher central charge} was introduced. This invariant was first defined in the mathematical literature~\cite{Ng_2019,Ng_2022}, and later given a physical interpretation as a criterion for the existence of a gapped boundary in~\cite{Kaidi_2022}, thereby providing a necessary and sufficient condition for the existence of a gapped boundary.

Its definition is as follows:
\begin{equation}
    \xi_n \coloneqq
    \frac{
        \sum\limits_{a \in \mathcal{A}} \theta(a)^n
    }{
        \bigl| \sum\limits_{a \in \mathcal{A}} \theta(a)^n \bigr|
    },
\end{equation}
where \( n \in \mathbb{N} \) satisfies
\begin{equation}
    \gcd\left(n, \frac{2|\mathcal{A}|}{\gcd(n, 2|\mathcal{A}|)}\right) = 1.
\end{equation}

The key result is the following:
\begin{lightgraybox}
An abelian TQFT admits a gapped boundary if and only if \( \xi_n \) is trivial for all \( n \in \mathbb{N} \) satisfying
\begin{equation}
    \gcd\left(n, \frac{2|\mathcal{A}|}{\gcd(n, 2|\mathcal{A}|)}\right) = 1.
\end{equation}
\end{lightgraybox}

Since
\begin{equation}
    \xi_n = \xi_{n + 2|\mathcal{A}|},
\end{equation}
only finitely many higher central charges need to be computed. These invariants thus provide a practical and computable alternative to the known criterion based on Lagrangian subgroups.
Later in this paper, we will denote
\begin{equation}
    \bN_c \coloneqq \left\{ n \in \bN \ \middle| \ 
    \gcd\left(n, \frac{2|\mathcal{A}|}{\gcd(n, 2|\mathcal{A}|)}\right) = 1
    \right\},\label{Def of Nc}
\end{equation}
That is, the higher central charge \( \xi_n \) is defined only for \( n \in \bN_c \).

    \section{Generalization of anomaly formula}
\subsection{Statement of formula}
Recall the anomaly formula \cite{Barkeshli:2016mew, Wang_2017, Orii_2025}:
\begin{equation}
    Z(\mathbb{RP}^4)\, Z(\mathbb{CP}^2) = \theta_{\mathcal{M}}. \label{ordinary anomaly formula}
\end{equation}
Here, both \( Z(\mathbb{RP}^4) \) and \( Z(\mathbb{CP}^2) \) take values in \( \{ \pm 1 \} \). In general, the time-reversal anomaly of 2+1-dimensional abelian bosonic systems is characterized by abelian bosonic SPT phases in 3+1-dimensional spacetime with time-reversal symmetry. These SPT phases are believed to be classified by \( \mathbb{Z}_2 \times \mathbb{Z}_2 \) \cite{kapustin2014symmetryprotectedtopologicalphases}, and can be distinguished by these two signs. In this sense, the formula~\eqref{ordinary anomaly formula} plays an important role, as it imposes a nontrivial constraint on the anomaly classification.

We can equivalently rewrite the left hand side of this equation as:
\begin{equation}
    \frac{\sum\limits_{a \in \Ker(1 - \mathsf{T})} \theta(a)\eta(a)}{\left| \sum\limits_{a \in \Ker(1 - \mathsf{T})} \theta(a)\eta(a) \right|}
    \cdot
    \frac{\sum\limits_{a \in \mathcal{A}} \theta(a)}{\left| \sum\limits_{a \in \mathcal{A}} \theta(a) \right|}.
\end{equation}
We now consider an analogue of the ordinary anomaly formula involving the higher central charge:
\begin{equation}
    \xi_n\coloneqq\frac{\sum\limits_{a \in \mathcal{A}} \theta(a)^n}
        {\left| \sum\limits_{a \in \mathcal{A}} \theta(a)^n \right|},
\end{equation}
and define
\begin{equation}
    \eta_n \coloneqq \frac{\sum\limits_{a \in \Ker(1 - \mathsf{T})} \big( \theta(a)\, \eta(a) \big)^n}{\left| \sum\limits_{a \in \Ker(1 - \mathsf{T})} \big( \theta(a)\, \eta(a) \big)^n \right|}.
\end{equation}

We then consider the product \( \eta_n\cdot\xi_n \) as the generalized anomaly formula. We find the following relation:
\begin{graybox}
    \begin{equation}
    \eta_n \cdot\xi_n = \frac{\sum\limits_{a \in \mathcal{E}^n} \theta(a)^n}{\left| \sum\limits_{a \in \mathcal{E}^n} \theta(a)^n \right|}.\label{generalized anomaly formula}
\end{equation}
\end{graybox}
Here, $\cE^n$ is defined by
\begin{equation}
       \cE^n\coloneqq\text{$\frac{\Ima^n(1+\sT)}{\Ima(1+\sT)}$-torsor}.
    \end{equation}
    satisfying
    \begin{equation}
        B(a,nb)=\eta(nb)\quad\text{for all }a\in\cE^n.
    \end{equation}

Before arriving at the final formula, we first define \( \mathcal{M}^n \) as follows:
\begin{equation}
\mathcal{M}^n \coloneqq \left\{ a \in \mathcal{A} \mid B(a, nb) = \eta(nb) \quad \text{for all } b \in \Ker(1 - \mathsf{T}) \right\}.
\end{equation}
We will first derive a formula involving \( \mathcal{M}^n \), and subsequently extract the crucial subset \( \mathcal{E}^n \) from it.

Here, \( n \) is taken from the set \( \bN_c \), which we introduced in~\ref{Def of Nc}. Note that this generalized anomaly formula, like \( \xi_n \), is defined for \( n \in \bN_c \).

\subsection{Derivation of formula}
As a preparation for the derivation, we will show the following three facts:
\begin{graybox}
    \begin{itemize}
        \item The quantity \( \eta_n \) takes the form
        \begin{equation}
            \eta_n = 
            \begin{cases}
                1 & \text{if } n \text{ is even}, \\
                Z(\mathbb{RP}^4) & \text{if } n \text{ is odd}.
            \end{cases}
            \label{generalized eta}
        \end{equation}
        \item 
\begin{equation}
    \left| \sum_{a \in \Ker(1 - \mathsf{T})} \big( \theta(a)\, \eta(a) \big)^n \right|=\begin{cases}
        |\Ker(1-\sT)|\quad &\text{if } n \text{ is even},\\
       |\cA|^{1/2} \quad &\text{if } n \text{ is odd}.
        
    \end{cases}\label{size of eta_n}
\end{equation}
        \item The quantity \( \widetilde{M}_{n,a} \), defined by
        \begin{equation}
            \widetilde{M}_{n,a} \coloneqq \sum_{b \in \Ker(1 - \mathsf{T})} B(-a, nb)\eta(nb),\label{def of tilda M}
        \end{equation}
        satisfies
        \begin{equation}
            \widetilde{M}_{n,a} = 
            \begin{cases}
                |\Ker(1 - \mathsf{T})| & \text{if } a \in \mathcal{M}^n, \\
                0 & \text{otherwise}.
            \end{cases}
            \label{tilda M}
        \end{equation}
    \end{itemize}
\end{graybox}
\paragraph{\eqref{generalized eta} :}
When $n$ is even, (say $n=2n'$):
\begin{equation}
    \big(\theta(a)\eta(a)\big)^{2n'}=\big(\theta(a)^2\eta(a)^2\big)^{n'}=\big(\theta(a)\theta(\sT a)\big)^{n'}=1.
\end{equation}
When $n$ is odd, (say $n=2n'+1$):
\begin{equation}
    \big(\theta(a)\eta(a)\big)^{2n'+1}=\big(\theta(a)\eta(a)\big)^{2n'}\theta(a)\eta(a)=\theta(a)\eta(a).
\end{equation}
Therefore, we conclude the proposition.

\paragraph{\eqref{size of eta_n} :}When \( n \) is even, the expression reduces to
\begin{equation}
    \left| \sum_{a \in \Ker(1 - \mathsf{T})} \big( \theta(a)\, \eta(a) \big)^n \right|
    = \sum_{a \in \Ker(1 - \mathsf{T})} 1
    = |\Ker(1 - \mathsf{T})|.
\end{equation}

When \( n \) is odd, it reduces to
\begin{equation}
    \left| \sum_{a \in \Ker(1 - \mathsf{T})} \big( \theta(a)\, \eta(a) \big)^n \right|
    = \left| \sum_{a \in \Ker(1 - \mathsf{T})} \theta(a)\, \eta(a) \right|
    = |\cA|^{1/2} \cdot |Z(\mathbb{RP}^4)|
    = |\cA|^{1/2}.
\end{equation}

\paragraph{\eqref{tilda M} :}We use the following fact:
\begin{lightgraybox}
    Define functions \( \rho^n_{B,a} \colon \Ker(1 - \mathsf{T}) \to U(1) \) and \( \rho^n_{\eta} \colon \Ker(1 - \mathsf{T}) \to \{\pm1\} \) by
        \begin{align*}
            \rho^n_{B,a}(b) &\coloneqq B(a, nb), \\
            \rho^n_{\eta}(b) &\coloneqq \eta(nb).
        \end{align*}
        Then, these are homomorphisms and thus define one-dimensional representations (i.e., characters) of the finite abelian group \( \Ker(1 - \mathsf{T}) \).
\end{lightgraybox}
We can rewrite the expression for \( \widetilde{M}_{n,a} \) as
\begin{equation}
    \widetilde{M}_{n,a} = \sum_{b \in \Ker(1 - \mathsf{T})} \overline{\rho^n_{B,a}(b)}\, \rho^n_{\eta}(b).
\end{equation}
Using the orthogonality of characters, we find:
\begin{equation}
    \widetilde{M}_{n,a} = 
    \begin{cases}
        |\Ker(1 - \mathsf{T})| & \text{if } \rho^n_{B,a} = \rho^n_{\eta}, \\
        0 & \text{otherwise},
    \end{cases}
    \label{amswer tilda M}
\end{equation}
which holds because the characters form an orthonormal basis under the standard pairing.

Recalling the definitions:
\begin{equation}
\rho^n_{B,a}(b) = B(a, nb), \qquad \rho^n_{\eta}(b) = \eta(nb),
\end{equation}
we see that \( \rho^n_{B,a} = \rho^n_{\eta} \) if and only if
\begin{equation}
B(a, nb) = \eta(nb) \quad \text{for all } b \in \Ker(1 - \mathsf{T}),
\end{equation}
i.e., \( a \in \mathcal{M}^n \). Therefore, we obtain:
\begin{equation}
    \widetilde{M}_{n,a} = 
    \begin{cases}
        |\Ker(1 - \mathsf{T})| & \text{if } a \in \mathcal{M}^n, \\
        0 & \text{otherwise}.
    \end{cases}
    \label{tilda M final}
\end{equation}

We now derive the following formula:
\begin{graybox}
    \begin{equation}
        \eta_n \cdot \xi_n = \frac{\sum\limits_{a \in \mathcal{M}^n} \theta(a)^n}{\left| \sum\limits_{a \in \mathcal{M}^n} \theta(a)^n \right|}.
    \end{equation}
\end{graybox}

To derive this, we use the following two facts shown above:
\begin{lightgraybox}
    \begin{itemize}
        \item 
      
\begin{equation}
    \left| \sum_{a \in \Ker(1 - \mathsf{T})} \big( \theta(a)\, \eta(a) \big)^n \right|=\begin{cases}
        |\Ker(1-\sT)|\quad &\text{if } n \text{ is even},\\
       |\cA|^{1/2} \quad &\text{if } n \text{ is odd}.
        
    \end{cases}\label{size of eta_n}
\end{equation}
        \item 
        \begin{equation}
            \widetilde{M}_{n,a} = 
            \begin{cases}
                |\Ker(1 - \mathsf{T})| & \text{if } a \in \mathcal{M}^n \\
                0 & \text{otherwise}.
            \end{cases}
        \end{equation}
    \end{itemize}
\end{lightgraybox}

We now compute \( \eta_n \cdot \xi_n \) explicitly:
\begin{equation}
    \begin{aligned}
        \eta_n \cdot \xi_n
        &= \frac{\sum\limits_{a \in \Ker(1 - \mathsf{T})} \big(\theta(a)\, \eta(a)\big)^n}
        {\left| \sum\limits_{a \in \Ker(1 - \mathsf{T})} \big(\theta(a)\, \eta(a)\big)^n \right|}
        \cdot
        \frac{\sum\limits_{b \in \mathcal{A}} \theta(b)^n}
        {\left| \sum\limits_{b \in \mathcal{A}} \theta(b)^n \right|} \\
        &= \frac{1}{|\Ker(1 - \mathsf{T})| \cdot \left| \sum\limits_{b \in \mathcal{A}} \theta(b)^n \right|}
        \sum_{a \in \Ker(1 - \mathsf{T})} \sum_{b \in \mathcal{A}} 
        \big( \theta(a)\, \eta(a)\, \theta(b) \big)^n.
    \end{aligned}
    \label{first to derive}
\end{equation}

Now observe:
\begin{equation}
    \begin{aligned}
        \sum_{a, b \in \mathcal{A}} \big( \theta(a)\, \eta(a)\, \theta(b) \big)^n
        &= \sum_{a, b \in \mathcal{A}} \big( \theta(a + b)\, B(-a, b)\, \eta(a) \big)^n \\
        &= \sum_{a, c \in \mathcal{A}} \big( \theta(c)\, B(-a, c - a)\, \eta(a) \big)^n \\
        &= \sum_{a, c \in \mathcal{A}} \big( \theta(c)\, B(a, a)\, B(-c, a)\, \eta(a) \big)^n.
    \end{aligned}
\end{equation}

Note that
\begin{equation}
    \begin{aligned}
        B(a,a)^n&=B(a,na)\\
        &=B(a,\sT na)\\
        &=B(a,\sT a)^n\\
        &=\eta\big(-(1+\sT)a\big)^n\\
        &=\eta\big(-n(1+\sT)a\big)
    \end{aligned}
\end{equation}
since \( a \in \Ker(1 - \mathsf{T}) \), i.e., \( \mathsf{T}na = na \).

Therefore, we obtain:
\begin{equation}
    \begin{aligned}
        \sum_{a, b \in \mathcal{A}} \big( \theta(a)\, \eta(a)\, \theta(b) \big)^n
        &= \sum_{a, b \in \mathcal{A}} \big( \theta(a + b)\, B(-a, b)\, \eta(a) \big)^n \\
        &= \sum_{a, c \in \mathcal{A}} \theta(c)^nB(-c,na)\eta(na)\eta\big(-n(1+\sT)a\big) \\
        &= \sum_{a, c \in \mathcal{A}} \theta(c)^nB(-c,na)\eta(-\sT na) \\
        &=\sum_{c\in\cA}\theta(c)^n\sum_{a\in\Ker(1-\sT)}B(-c,na)\eta(na)\\
        &=\sum_{c\in\cA}\theta(c)^n\cdot \widetilde{M}_{n,c}\\
        &=|\Ker(1-\sT)|\sum_{c\in\cM^n}\theta(c)^n
    \end{aligned}
    \label{second to derive}
\end{equation}
where we used the explicit form of \( \widetilde{M}_{n,c} \) from~\eqref{tilda M final}.

Substituting this back into~\eqref{first to derive}, we conclude:
\begin{equation}
    \eta_n \cdot \xi_n
        = \frac{|\Ker(1 - \mathsf{T})|}{\left| \sum\limits_{a \in \Ker(1 - \mathsf{T})} \big(\theta(a)\, \eta(a)\big)^n \right| \cdot \left| \sum\limits_{b \in \mathcal{A}} \theta(b)^n \right|} 
        \cdot  \sum_{c \in \mathcal{M}^n} \theta(c)^n .\label{deriving formula}
\end{equation}
Note that we have
\begin{equation}
    \left| \sum_{a \in \Ker(1 - \mathsf{T})} \big( \theta(a)\, \eta(a) \big)^n \right|=\begin{cases}
        |\Ker(1-\sT)|\quad &\text{if } n \text{ is even},\\
       |\cA|^{1/2} \quad &\text{if } n \text{ is odd}.
        
    \end{cases}\label{size of eta_n}
\end{equation}
using this equation~\eqref{deriving formula}, we obtain
\begin{equation}
    \eta_n \cdot \xi_n = \begin{cases}
    \frac{1}{\left| \sum\limits_{a \in \mathcal{A}} \theta(a)^n \right|} 
        \cdot  \sum_{a\in \mathcal{M}^n} \theta(a)^n \quad &\text{if } n \text{ is even},\\
       \frac{\left|\mathrm{Ker}(1-\sT) \right|}{|\cA|^{1/2}\cdot \left| \sum\limits_{a \in \mathcal{A}} \theta(a)^n \right|} 
        \cdot  \sum_{a \in \mathcal{M}^n} \theta(a)^n\quad &\text{if } n \text{ is odd}.
    \end{cases}
\end{equation}
Now note that \(|\eta_n \cdot \xi_n |=1\), we obtain
\begin{equation}
    \left|\sum_{a \in \mathcal{M}^n} \theta(a)^n\right|=\begin{cases}
    \left| \sum\limits_{a \in \mathcal{A}} \theta(a)^n \right|\quad&\text{if } n \text{ is even},\\
    \frac{|\cA|^{1/2}\cdot \left| \sum\limits_{a \in \mathcal{A}} \theta(a)^n \right|}{\left|\mathrm{Ker}(1-\sT)\right|} \quad&\text{if } n \text{ is odd}.
\end{cases}
\end{equation}
Therefore,
\begin{equation}
        \eta_n \cdot \xi_n = \frac{\sum\limits_{a \in \mathcal{M}^n} \theta(a)^n}{\left| \sum\limits_{a \in \mathcal{M}^n} \theta(a)^n \right|}
    \end{equation}
This completes half of the derivation.\footnote{This entire derivation is completely analogous to the discussion in \cite{Orii_2025}.}

In the next subsections, we specify the structure of $\cM^n$ and reduce the 
summention over $\cM^n$ to one over $\cE^n$.

\subsection{Insight from Dimensions}
\label{subsec.dim}

Before proceeding to analyze the detailed structure of \( \mathcal{M}^n \), let us first consider its meaning from the perspective of Hilbert space dimensions. As shown in~\cite{Barkeshli:2016mew}, particularly in equation~(143), the dimension of the Hilbert space on \( \Sigma_n(a_1, \dots, a_k) \) is given by
\begin{equation}
    \dim\left( V\big( \Sigma_n(a_1, \dots, a_k) \big) \right)
    = \frac{1}{|\mathcal{A}|^{\frac{2 - k}{2}}} \sum_{b \in \Ker(1 - \mathsf{T})} B\big( - (a_1 + \cdots + a_k), b \big)\, \eta(n b),
\end{equation}
where \( \Sigma_n(a_1, \dots, a_k) \) denotes a non-orientable surface with \( n \) crosscaps and \( k \) punctures labeled by \( a_1, \dots, a_k \).

If we impose the constraint
\begin{equation}
a_1 + \cdots + a_k = n a,
\end{equation}
then the formula becomes
\begin{equation}
    \dim\left( V\big( \Sigma_n(a_1, \dots, a_k) \big) \right)
    = \frac{1}{|\mathcal{A}|^{\frac{2 - k}{2}}} \sum_{b \in \Ker(1 - \mathsf{T})} B(-a, nb)\, \eta(n b)
    = \frac{1}{|\mathcal{A}|^{\frac{2 - k}{2}}} \widetilde{M}_{n,a}.
\end{equation}
From this point of view, we can identify \( \mathcal{M}^n \) as the subset
\begin{graybox}
\begin{equation}
    \mathcal{M}^n = \left\{ a \in \mathcal{A} \;\middle|\; \dim\left( V\big( \Sigma_n(na) \big) \right) \ne 0 \right\}.
\end{equation}
\end{graybox}

In fact, we can generalize this statement slightly. Let us define \( \Sigma_{n,m}(a_1, \dots, a_k) \) as a non-orientable surface with \( m \) crosscaps and \( k \) punctures labeled by \( a_1, \dots, a_k \), subject to the constraints:
\begin{equation}
    \begin{aligned}
        a_1 + \cdots + a_k &= n a, \\
        n &\equiv m \mod 2.
    \end{aligned}
\end{equation}
Then, since \( \eta(a) \in \{\pm1\} \), it follows that
\begin{equation}
    \left\{ a \in \mathcal{A} \;\middle|\; \dim\left( V\big( \Sigma_n(a_1, \dots, a_k) \big) \right) \ne 0 \right\}
    = \left\{ a \in \mathcal{A} \;\middle|\; \dim\left( V\big( \Sigma_{n,m}(a_1, \dots, a_k) \big) \right) \ne 0 \right\}.
    \label{mod 2 reduction}
\end{equation}

\paragraph{Recall when \( n = 1 \):} In our previous work~\cite{Orii_2025}, we proved that
\begin{equation}
    \dim\left( V\big( \Sigma_1(a) \big) \right) = \dim\left( V\big( \mathbb{RP}^2(a) \big) \right) \ne 0
\end{equation}
for all \( a \in \mathcal{M} \), and established several results that we reviewed in Section~3. In this sense, the set \( \mathcal{M}^n \) naturally generalizes \( \mathcal{M} \) from the perspective of Hilbert space dimension as well.

\paragraph{The fact modulo 2:} The result stated in~\eqref{mod 2 reduction} has a geometric interpretation. By Dyck's Theorem~\cite{Francis01051999}, two crosscaps can be replaced with a handle, provided that the overall non-orientability of the surface is preserved. For example,
\begin{equation}
    \begin{aligned}
        \mathbb{RP}^2 \# \mathbb{RP}^2 \# \mathbb{RP}^2 &\cong \mathbb{RP}^2 \# T^2, \\
        \mathbb{RP}^2 \# \mathbb{RP}^2 \# \mathbb{RP}^2 \# \mathbb{RP}^2 &\cong \mathbb{RP}^2 \# \mathbb{RP}^2 \# T^2 \cong \mathbb{KB} \# T^2,
    \end{aligned}
\end{equation}
where \( T^2 \) is the torus and \( \mathbb{KB} \) denotes the Klein bottle. Therefore, it is quite natural that the positivity of the Hilbert space dimension depends on the number of crosscaps modulo 2.

\subsection{Structure of $\cM^n$}

\paragraph{Expression of \( \cM^n \):}\label{para:expression of M^n}Now, let us define $\Ima^n(1+\sT)$ 
as
\begin{equation}
    \Ima^n(1+\sT)\coloneqq\{a\in\cA\mid na\in\Ima(1+\sT)\}.
\end{equation}
Then we find that
\begin{graybox}
the expression of $\cM^n$ is
   \begin{equation}
\cM^n\coloneqq\text{$\Ima^n(1+\sT)$-torsor}.
   \end{equation}
   satisfying, for all $a\in\cM^n$
\begin{equation}
    B(a,nb)=\eta(nb)\quad\text{for all }b\in\Ker(1-\sT).
\end{equation}
\end{graybox}

To prove this structure, we verify the following:
    \begin{itemize}
        \item If \( a \in \cM^n \) and \( b \in \Ima^n(1 + \sT) \) satisfy \( b + a = a \), then \( b = 0 \).
        \item For any \( a, b \in \cM^n \), there exists \( c \in \Ima^n(1 + \sT) \) such that \( a = b + c \), i.e., \( c = a - b \).
    \end{itemize}
The first condition is trivial. For the second, we proceed as follows:

Assume \( a, b \in \cM^n \), so both satisfy
\begin{equation}
    \begin{cases}
        B(a, nd) = \eta(nd) \\
        B(b, nd) = \eta(nd)
    \end{cases}
    \quad \text{for all } d \in \Ker(1 - \sT).
\end{equation}
Dividing the two equations gives:
\begin{equation}
    B(a - b, nd) = B(n(a - b), d) = 1 \quad \text{for all } d \in \Ker(1 - \sT).
\end{equation}
By non-degeneracy of \( B \), this implies:
\begin{equation}
    n(a - b) \in \Ker(1 - \sT)^\perp = \Ima(1 + \sT),
\end{equation}
so we define \( c := a - b \in \Ima^n(1 + \sT) \), and get the conclusion as desired.


\paragraph{Constraint on addition :}\label{para:addition}
From the fact that \( \cM^n\) forms a torsor over \( \Ima^n(1 + \sT) \), we can easily derive the following constraint:
\begin{graybox}
    \begin{equation}
        a + b \in \Ima^n(1 + \sT) \quad \text{for all } a, b \in \cM^n.
    \end{equation}
\end{graybox}
It suffices to show that \( 2a \in \Ima^n(1 + \sT) \), since the torsor property implies that any difference \( a - b \in \Ima^n(1 + \sT) \), and hence the sum \( a + b = 2a + (b - a) \in \Ima^n(1 + \sT) \) if \( 2a \in \Ima^n(1 + \sT) \).

To prove this, let \( a \in \cM^n \). Then for any \( b \in \Ker(1 - \sT) \), we have:
\begin{equation}
    B(a, nb)^2 = \eta(nb)^2 = 1,
\end{equation}
since \( \eta(nb) \in \{ \pm 1 \} \). On the other hand, using bilinearity of \( B \), the left-hand side simplifies as:
\begin{equation}
    B(a, nb)^2 = B(2a, nb) = B(2na, b).
\end{equation}
Therefore,
\begin{equation}
    B(2na, b) = 1 \quad \text{for all } b \in \Ker(1 - \sT).
\end{equation}
By non-degeneracy of \( B \), this implies:
\begin{equation}
    2na \in \Ker(1 - \sT)^\perp = \Ima(1 + \sT),
\end{equation}
i.e.,
\begin{equation}
    2a\in\Ima^n(1 + \sT)
\end{equation}
as desired.

\paragraph{Constraint on $\sT a$ :}We can derive another proposition as follows:
\begin{graybox}
    \begin{equation}
        \sT a\in\cM^n\quad\text{for all }
        a\in\cM^n.\label{tna}
    \end{equation}
\end{graybox}
This follows easily since
\begin{equation}
    \begin{aligned}
        B(\sT a,nb)&=B(a,-\sT nb)\\
        &=\eta(-\sT nb)\\
        &=\eta(nb)\quad\text{for all }b\in\Ker(1-\sT).
    \end{aligned}
\end{equation}

\subsection{Topological spins of $\cM^n$}\label{subsec:top spin}
To fully understand the structure of $\cM^n$, let us define the following:
\begin{graybox}
    Define 
    \begin{equation}
       \cE^n\coloneqq\text{$\frac{\Ima^n(1+\sT)}{\Ima(1+\sT)}$-torsor}.
    \end{equation}
    satisfying
    \begin{equation}
        B(a,nb)=\eta(nb)\quad\text{for all }a\in\cE^n.
    \end{equation}
    then, we find that
    \begin{equation}
    \text{$\theta(a)^n$ is well defined on $\cE^n$.}\label{constancy of theta^n}
\end{equation}
\end{graybox}

We will use the following identities:

\begin{lightgraybox}
\begin{itemize}
    \item The compatibility condition between \( B \) and \( \eta \) (see Appendix~\ref{App2}):
    \begin{equation}
    B(a, \sT a)\, \eta((1 + \sT)a) = 1 \quad \text{for all } a \in \cA.
    \end{equation}
    
    \item Anti-unitarity of \( \sT \):
    \begin{equation}
    \theta(a)\, \theta(\sT a) = 1 \quad \text{for all } a \in \cA.
    \end{equation}
    
    \item Since \( a \in \cM^n \), we also have:
    \begin{equation}
    \eta(-n(1 + \sT)c)\, B(a, n(1 + \sT)c) = 1\quad \text{for all } c \in \cA.
    \end{equation}
\end{itemize}
\end{lightgraybox}

Using these, we compute:
\begin{equation}
\begin{aligned}
    \frac{\theta(b)^n}{\theta(a)^n}
    &= \left( \frac{\theta(a + (1 + \sT)c)}{\theta(a)} \right)^n \\
    &= \left( \frac{\theta(a)\, \theta((1 + \sT)c)\, B(a, (1 + \sT)c)}{\theta(a)} \right)^n \\
    &= \left( \theta((1 + \sT)c)\, B(a, (1 + \sT)c) \right)^n \\
    &= \left( \theta(c)\, \theta(\sT c)\, B(c, \sT c)\, B(a, (1 + \sT)c) \right)^n \\
    &= \left( \eta(-(1 + \sT)c)\, B(a, (1 + \sT)c) \right)^n \\
    &= \eta(-n(1 + \sT)c)\, B(a, n(1 + \sT)c) \\
    &= 1.
\end{aligned}
\end{equation}

Therefore, we conclude the desired identity holds. 

Furthermore, we can constrain the possible values of \( \theta(a)^n \) as follows:

\begin{graybox}
    \begin{equation}
        \theta(a)^n \in \{ \pm 1 \}\quad\text{for all }a\in\cE^n. \label{pm1 theta mn}
    \end{equation}
\end{graybox}

Combining equations~\eqref{tna} and~\eqref{constancy of theta^n}, we obtain
\begin{equation}
    \theta(a)^n = \theta\big(a-(1+\sT)a\big)^n = \theta(-\sT a)^n = \theta(\sT a)^n=\overline{\theta(a)^n},
\end{equation}
which implies
\begin{equation}
    \left( \theta(a)^n \right)^2 = 1.
\end{equation}
Therefore, the claim follows.

\subsection{Anomaly formula and higher central charge}
Let us recall the expression of the formula and the well-defininedness of $\theta(a)^n$. Then the generalized anomaly formula simplifies to the following expression:
\begin{graybox}
    \begin{equation}
        \eta_n \cdot \xi_n = \frac{\sum\limits_{a \in \cE^n} \theta(a)^n}{\left| \sum\limits_{a \in \cE^n} \theta(a)^n \right|}.
     \label{reduction of formula}
    \end{equation}
\end{graybox}

From the expression above and equation~\eqref{generalized eta}, we obtain the following formula:
\begin{equation}
    \xi_n = 
    \begin{cases}
        \dfrac{\sum\limits_{a \in \cE^n} \theta(a)^n}{\left| \sum\limits_{a \in \cE^n} \theta(a)^n \right|} & \text{if } n \text{ is even}, \\
        Z(\mathbb{RP}^4) \cdot \dfrac{\sum\limits_{a \in \cE^n} \theta(a)^n}{\left| \sum\limits_{a \in \cE^n} \theta(a)^n \right|} & \text{if } n \text{ is odd},
    \end{cases}
\end{equation}

The comparison table between the original formula and the generalized one is shown in Table~\ref{tab:Mn-summary}.

\begin{table}[htbp]
\centering
\small 
\renewcommand{\arraystretch}{2.5} 

\begin{tabular}{|>{\centering\arraybackslash}m{2.1cm}
                |>{\centering\arraybackslash}m{6.6cm}
                |>{\centering\arraybackslash}m{6.6cm}|}
  \hline
  & \( n = 1 \) & \( n \in \mathbb{N}_c \) \\
  \hline
  \textbf{Definition} &
  \(
    \left\{
      \begin{array}{c}
        a \in \cA \quad\text{s.t. } B(a,b) = \eta(b) \\
        \forall b \in \Ker(1 - \sT)
      \end{array}
    \right\}\) &
  \(
    \left\{
      \begin{array}{c}
        a \in \cA \quad\text{s.t. } B(a,nb) = \eta(nb) \\
        \forall b \in \Ker(1 - \sT)
      \end{array}
    \right\}\) \\
  \hline
  \textbf{Structure} &
  Torsor over \( \Ima(1 + \sT) \) &
  Torsor over \( \Ima^n(1 + \sT) \) \\
  \hline
  \textbf{Topological spin} &
  \( \theta(a)  \in \{\pm 1\} \quad \forall a \in \cM \) &
  \( \theta(a)^n \in \{\pm 1\} \quad \forall a \in \cM^n \) \\
  \hline
  \textbf{Dimension} &
  \( \dim\left( V\big( \Sigma_1(a) \big) \right) \ne 0 \quad\forall a\in\cM\) &
  \( \dim\left( V\big( \Sigma_n(na) \big) \right) \ne 0 \quad\forall a\in\cM^n\) \\
  \hline
  \textbf{Addition} &
  \( 2a \in \Ima(1 + \sT) \quad\forall a\in\cM\) &
  \( 2a\in \Ima^n(1 + \sT)\quad\forall a\in\cM^n \) \\
  \hline
  \textbf{Formula} &
  \( Z(\mathbb{RP}^4) \cdot Z(\mathbb{CP}^2) = \theta_{\mathcal{M}} \) &
  \( \eta_n \cdot \xi_n = \frac{\sum\limits_{a \in \cE^n} \theta(a)^n}{\left| \sum\limits_{a \in \cE^n} \theta(a)^n \right|} \) \\
  \hline
  \textbf{Meaning of formula} &
  Constraint on anomaly &
  \textbf{\Large ?} \\
  \hline
\end{tabular}
\caption{Comparison of the structures of \( \cM \) and \( \cM^n \) under time-reversal symmetry.}
\label{tab:Mn-summary}
\end{table}
    \section{Conclusion and outlook}
In this paper, we present a generalization of the anomaly formula incorporating time-reversal symmetry, which naturally accounts for the higher central charge. The resulting formula and its structure closely resemble the known original case, yet its direct physical interpretation remains unclear. Given that $\mathcal{M}^n$ can be viewed as a subsystem that nontrivially contributes to the higher central charge, it is plausible that this subset plays an essential role in determining the existence of a gapped boundary.

As future directions, it would be valuable to elucidate the physical interpretation of the generalized formula and to extend the framework to non-abelian systems and spin-TQFTs.

\paragraph{Acknowledgments.}
The author would like to thank Y.~Tachikawa for many fruitful discussions. He is also grateful to K.~Ohmori and R.~Kobayashi for providing helpful insights related to this work and its possible extensions. In addition, the author thanks T.~Ando; conversations with him motivated the consideration of higher central charges, which in turn led to the formulation of the generalized anomaly formula. Lastly, he would like to thank the anonymous referee for their helpful comments.

This research is supported in part by the Forefront Physics and Mathematics Program to Drive Transformation (FoPM), a World-leading Innovative Graduate Study (WINGS) Program at the University of Tokyo.

    \appendix
\section{Orthogonality of $\Ker(1-\sT)$ and $\Ima(1+\sT)$}\label{App 1}
In this section, we will prove the following equalities:
\begin{graybox}
    \begin{equation}
    \Ker(1 - \sT) = \left[ \Ima(1 + \sT) \right]^\perp, \quad 
    \Ker(1 + \sT) = \left[ \Ima(1 - \sT) \right]^\perp.
\end{equation}
\end{graybox}
This result was shown in~\cite{Wang_2017, Lee:2018eqa}. For the reader's convenience, we briefly review the derivation here.

We first note the basic identity:
\begin{equation}
    B(\sT a, b) = B(a, \sT b)^{-1}. \label{eq:B-inverse}
\end{equation}
This implies the following symmetry relation:
\begin{equation}
    B((1 + \sT)a, b) = B(a, (1 - \sT)b). \label{eq:B-twisted}
\end{equation}

As a consequence, we obtain the inclusions:
\begin{equation}
    \Ker(1 - \sT) \subset \left[ \Ima(1 + \sT) \right]^\perp, \quad 
    \Ker(1 + \sT) \subset \left[ \Ima(1 - \sT) \right]^\perp. \label{eq:kernel-subset}
\end{equation}

Using the non-degeneracy of the bilinear form $B$, we obtain the inequalities:
\begin{equation}
    |\Ker(1 - \sT)| \le \frac{|\cA|}{|\Ima(1 + \sT)|}, \quad
    |\Ker(1 + \sT)| \le \frac{|\cA|}{|\Ima(1 - \sT)|}. \label{eq:kernel-bound}
\end{equation}

On the other hand, it is evident that
\begin{equation}
    |\cA / \Ker(1 + \sT)| = |\Ima(1 + \sT)|, \quad 
    |\cA / \Ker(1 - \sT)| = |\Ima(1 - \sT)|. \label{eq:image-vs-kernel}
\end{equation}

Combining equations~\eqref{eq:kernel-bound} and~\eqref{eq:image-vs-kernel}, we conclude that the inclusions in~\eqref{eq:kernel-subset} are in fact equalities:
\begin{equation}
    \Ker(1 - \sT) = \left[ \Ima(1 + \sT) \right]^\perp, \quad 
    \Ker(1 + \sT) = \left[ \Ima(1 - \sT) \right]^\perp. \label{eq:orthogonal-identities}
\end{equation}

\section{Consistency condition for $B$ and $\eta$}\label{App2}
In this section, we derive the following identity:
\begin{graybox}
\begin{equation}
    B(a, \sT a)\, \eta\big((1 + \sT)a\big) = 1 \quad \text{for all } a \in \cA.
\end{equation}
\end{graybox}

This identity is regarded as a consistency condition that must be satisfied by the bilinear form $B$ and the homomorphism $\eta$, as discussed in~\cite[Eq.~(34)]{Barkeshli:2017rzd}, \cite[Sec.~2.4]{Lee:2018eqa}, and \cite[Sec.~4.3]{Orii_2025}.

We derive this condition from the following property:
\begin{lightgraybox}
For all $a \in \cM$, where $\cM$ is an $\Ima(1 + \sT)$-torsor, the following hold:
\begin{equation}
\begin{aligned}
    \theta(a) &= \text{const.}, \\
    B(a, b) &= \eta(b) \quad \text{for all } b \in \Ker(1 - \sT).
\end{aligned}
\end{equation}
\end{lightgraybox}

Using these properties, we compute:
\begin{equation}
\begin{aligned}
    1 &= \frac{\theta(a + (1 + \sT)c)}{\theta(a)} \\
      &= \frac{\theta(a)\, \theta((1 + \sT)c)\, B(a, (1 + \sT)c)}{\theta(a)} \\
      &= \theta(c)\, \theta(\sT c)\, B(c, \sT c)\, \eta((1 + \sT)c) \\
      &= B(c, \sT c)\, \eta((1 + \sT)c),
\end{aligned}
\end{equation}
which holds for all $a \in \cM$ and $c \in \cA$.

This completes the derivation.


\bibliographystyle{ytamsalpha}
 \def\arxivfont{\rm}
 \baselineskip=.95\baselineskip
\bibliography{ref}

\end{document}